\documentclass[notitlepage,prd,twocolumn,showpacs,preprintnumbers,nofootinbib,superscriptaddress,floatfix,tightenlines, longbibliography]{revtex4-1}
\usepackage{mathrsfs}
\usepackage{amssymb}
\usepackage{amsmath}
\usepackage{graphicx}
\usepackage[utf8]{inputenc}
\usepackage{amsfonts,amsbsy}
\usepackage{nicefrac}
\usepackage{xcolor}
\definecolor{lcolor}{rgb}{0.5,0,0}
\definecolor{citcolor}{rgb}{0,0.3,0.0}
\usepackage[breaklinks,colorlinks,urlcolor=blue,citecolor=citcolor,linkcolor=lcolor]{hyperref}
\usepackage{comment}
\usepackage{bbm}
\usepackage[normalem]{ulem} 
\usepackage{enumitem}

\newcommand{\qhat}{\hat{q}}

\newcommand{\qs}{Q_\mathrm{s}}

\newcommand{\fig}{Fig.~}

\newcommand{\eq}{Eq.~}
\newcommand{\eqs}{Eqs.~}

\newcommand{\re}{Ref.~}
\newcommand{\res}{Refs.~}
\newcommand{\app}{App.~}

\newcommand{\bs}[1]{\boldsymbol{#1}}

\newcommand{\ud}{\mathrm{d}}

\newcommand{\nn}{\nonumber}

\newcommand{\nc}{N_c}

\newcommand{\der}{\mathrm{d}}

\newcommand{\be}{\begin{equation}}
\newcommand{\ee}{\end{equation}}
\newcommand{\bea}{\begin{eqnarray}}
\newcommand{\eea}{\end{eqnarray}}

\newcommand{\derthree}{\dfrac{\der^3 p }{\left(2 \pi \right)^3}}

\newcommand{\tauT}{\tau_{\mathrm{BMSS}}}
\newcommand{\tauR}{\tau_R}
\newcommand{\Teps}{T_{\varepsilon}}
\newcommand{\pxi}{p_\xi}

\begin{document}

\title{Limiting attractors in heavy-ion collisions}

\author{K.~Boguslavski} 
\affiliation{Institute for Theoretical Physics, TU Wien, Wiedner Hauptstra{\ss}e 8-10, 1040 Vienna, Austria}

\author{A.~Kurkela} 
\affiliation{Faculty of Science and Technology, University of Stavanger, 4036 Stavanger, Norway}

\author{T.~Lappi} 
\affiliation{Department of Physics, P.O.~Box 35, 40014 University of Jyv\"{a}skyl\"{a}, Finland}
\affiliation{Helsinki Institute of Physics, P.O.~Box 64, 00014 University of Helsinki, Finland}

\author{F.~Lindenbauer} 
\affiliation{Institute for Theoretical Physics, TU Wien, Wiedner Hauptstra{\ss}e 8-10, 1040 Vienna, Austria}

\author{J.~Peuron} 
\affiliation{Department of Physics, P.O.~Box 35, 40014 University of Jyv\"{a}skyl\"{a}, Finland}
\affiliation{Helsinki Institute of Physics, P.O.~Box 64, 00014 University of Helsinki, Finland}
\affiliation{Dept.   of  Astronomy  and  Theoretical  Physics,  S\"{o}lvegatan  14A,  S-223  62  Lund,  Sweden}

\begin{abstract}
    We study universal features of the hydrodynamization process in heavy-ion collisions using QCD kinetic theory simulations for a wide range of couplings. We introduce the new concept of limiting attractors, which are obtained by extrapolation to vanishing and strong couplings. While the hydrodynamic limiting attractor emerges at strong couplings and is governed by the viscosity-related relaxation time scale $\tauR$, we identify a bottom-up limiting attractor at weak couplings. It corresponds to the late stages of the perturbative bottom-up thermalization scenario and exhibits isotropization on the time scale $\tau_{\text{BMSS}} = \alpha_s^{-13/5}/Q_s$. In contrast to hydrodynamic limiting attractors, at finite couplings the bottom-up limiting attractor provides a good universal description of the pre-hydrodynamic evolution of jet and heavy-quark momentum broadening ratios $\qhat^{yy} / \qhat^{zz}$ and $\kappa_T/\kappa_z$. We also provide parametrizations for these ratios for phenomenological studies of pre-equilibrium effects on jets and heavy quarks.
\end{abstract}

\maketitle


\section{Introduction}

Universal features that emerge during the pre-equilibrium evolution of the quark-gluon plasma (QGP) generated in relativistic heavy-ion collisions have attracted a lot of attention \cite{Kurkela:2011ub, Berges:2014bba, Berges:2013eia, Blaizot:2017ucy, Kurkela:2019set}. In recent years, the development of the concept of nonequilibrium attractors \cite{Heller:2015dha, Blaizot:2017ucy, Kurkela:2019set, Almaalol:2020rnu, Du:2022bel} (for reviews see \cite{Jankowski:2023fdz, Soloviev:2021lhs})
has significantly deepened our understanding of how the system reaches isotropy and thermal equilibrium.
An attractor refers to the property that, for various initial conditions, the time evolution of the system at sufficiently late times follows a universal curve that is characterized by a reduced number of parameters \cite{Du:2022bel}. 
This property allows making predictions from the pre-equilibrium stages \cite{Giacalone:2019ldn, Garcia-Montero:2023lrd} despite incomplete information about the initial conditions. Furthermore, it has been observed across several different models and values of the coupling constant that the attractors seem to be functions of just a single rescaled time variable. This time variable is typically given in units of a relaxation time scale $\tauR$,  which is a combination of the temperature and the shear viscosity over entropy density ratio $\eta/s$.
This scaling has been observed in various formulations of hydrodynamics \cite{Strickland:2017kux,Kurkela:2019set}, quasi-particle models described by kinetic theory at weak couplings \cite{Baier:2000sb, Arnold:2002zm, AbraaoYork:2014hbk, Kurkela:2015qoa, Keegan:2015avk, Kurkela:2018xxd, Du:2020zqg}, and in holographic models at strong coupling \cite{Heller:2012km, Keegan:2015avk, Heller:2016gbp, Kurkela:2019set}. 

While kinetic theory naturally encompasses hydrodynamics \cite{Denicol:2012cn, Baier:2007ix}, a kinetic theory description of the medium possesses a richer structure due to a larger set of degrees of freedom. An analysis of QCD in the weak-coupling limit where it is described by an effective kinetic theory \cite{Arnold:2002zm} reveals a bottom-up picture of thermalization that undergoes several dynamical stages \cite{Baier:2000sb, Kurkela:2015qoa, Kurkela:2018xxd, Du:2020zqg}. In this scenario, thermalization is reached on a time scale $\qs \tauT \sim \alpha_s^{-13/5}$, where $\alpha_s$ is the strong coupling constant and $\qs$ is the initial typical momentum scale of hard excitations. 

The thermalization times $\tauR$ and $\tauT$ become parametrically different for small couplings. This motivates us to study how the two isotropization pictures based on the bottom-up scenario and a hydrodynamic attractor are connected to each other. We will demonstrate that the hydrodynamic attractor can be reinterpreted as a \emph{limiting attractor} in an extrapolation to infinite coupling. Additionally, we find a bottom-up limiting attractor, obtained by extrapolating to vanishing coupling. 

While for phenomenologically relevant values of the coupling $\lambda \gtrsim 10$ the $\tauR$-scaling has been observed in several important bulk observables, this does not need to be the case for hard probes like jets and heavy quarks. These may be dominated by different sectors of the momentum distribution, making them potentially more sensitive to the large scale separations present in the pre-equilibrium bottom-up scenario. Motivated by this, we study the heavy quark momentum diffusion coefficient $\kappa$ \cite{Moore:2004tg} and the jet quenching parameter $\qhat$ \cite{Baier:1996sk}, which encode medium effects to hard probes. The behavior of these coefficients out of equilibrium has recently attracted increased attention \cite{Ipp:2020nfu, Ipp:2020mjc, Boguslavski:2020tqz, Carrington:2020sww, Hauksson:2021okc, Khowal:2021zoo, Carrington:2021dvw, Carrington:2022bnv, Ruggieri:2022kxv, Hauksson:2023tze, Avramescu:2023qvv, Boguslavski:2023alu, Boguslavski:2023fdm, Prakash:2023wbs, Du:2023izb}.

By studying the scaling of various observables using QCD effective kinetic theory, we find that at weak coupling, the far-from-equilibrium behavior leading to isotropization exhibits $\tauT$-scaling associated with the bottom-up limiting attractor while the final approach at very late times is governed by $\tauR$-scaling. With increasing coupling, the scale separations of the bottom-up scenario weaken and the hydrodynamic limiting attractor describes an increasingly large part of the full non-equilibrium evolution down to earlier times and farther away from equilibrium.  

Moreover, we also apply these attractors to anisotropy ratios of $\kappa$ and $\hat q$ that measure differences in the momentum broadening of heavy quarks and jets. We find that the hydrodynamic limiting attractor for these observables gives a less accurate description of the finite-coupling results. In contrast, the bottom-up limiting attractor starts describing the data at much earlier times and should be considered when performing phenomenological modeling of heavy quarks and jets in heavy-ion collisions.

\section{Theory and setup}
\label{sec:theory}

\subsection{Effective kinetic theory}

We run simulations using the effective kinetic theory of \re\cite{Arnold:2002zm} with the same setup as in our previous works \cite{Boguslavski:2023alu, Boguslavski:2023fdm}. Here the pre-equilibrium plasma is described in terms of gluons, represented by their quasiparticle distribution function $f(\bs{p})$.
These gluons are the dominant degree of freedom before chemical equilibration, which takes place after hydrodynamization \cite{Kurkela:2018xxd}.
The  time evolution is given by the Boltzmann equation in proper time~$\tau$
\begin{equation}
\label{eq:EKTEOMS}
-\dfrac{\partial f(\bs{p})}{\partial \tau } = \mathcal{C}_{1 \leftrightarrow 2  }[f(\bs{p})] + \mathcal{C}_{2 \leftrightarrow 2  }[f(\bs{p})] + \mathcal{C}_{\mathrm{exp} }[f(\bs{p})].
\end{equation}
Here $\mathcal{C}_{1 \leftrightarrow 2  }$ encodes effective particle 1 to 2 splittings and 2--to--2 scatterings are described by $\mathcal{C}_{2 \leftrightarrow 2  }$. The expansion is included in the effective expansion term \cite{Mueller:1999pi}, which in the boost invariant case takes the simple form $\mathcal{C}_{\mathrm{exp} }[f(\bs{p})] = - \frac{p_z}{\tau} \frac{\partial}{\partial p_z} f(p)$. We assume that our distribution depends on the magnitude of the momentum $p$ and on the polar angle $\theta$ with $\cos\theta = \hat{\bs{z}} \cdot \hat{\bs{p}}$, i.e $f(\bs{p}) = f(p, \cos \theta_p)$ but not on the azimuthal angle $\phi$. 

We will need the energy-momentum tensor that is calculated in kinetic theory as 
\begin{equation}
T^{\mu \nu} = \nu_g \int \derthree \dfrac{p^\mu p^\nu }{p} f(\bs p).
\end{equation}
Here, $\nu_g=2(\nc^2-1)$ counts the number of degrees of freedom for gluons.
The longitudinal and transverse pressures are defined as $P_T = T_{xx}=T_{yy}$ and $P_L = T_{zz}$. The typical occupancy of hard particles is given by 
\begin{equation}
\label{eq:occupancy}
\frac{\langle \lambda f p\rangle}{\langle p \rangle} = \lambda\,\frac{\int \ud^3 p\,p\,f^2}{\int \ud^3 p\,p\,f}.
\end{equation}
For a more detailed description of effective kinetic theory and its implementation and discretization details, we refer to Refs.~\cite{Arnold:2002zm, Kurkela:2015qoa}. 

In the next subsections, we briefly describe how we extract the diffusion coefficient and jet quenching parameter out of equilibrium from our kinetic theory simulations.

\subsection{Heavy quark momentum diffusion coefficient}

At leading order, the heavy quark momentum diffusion coefficient $\kappa$ for pure glue QCD is given by \cite{Moore:2004tg}
\begin{align}
\label{eq:kappa_master_formula}
\kappa_i &= \frac{1}{2 M}\int_{\bs{k} \bs{k^\prime} \bs{p^\prime}} \left(2 \pi \right)^3 \delta^3\left( \bs{p} +\bs{k} - \bs{p^\prime} - \bs{k^\prime} \right) \nn \\
& \times  2 \pi \delta \left(k^\prime - k \right) q_i^2 
 \left| \mathcal{M}_g \right|^2 f(\bs{k}) (1+f(\bs{k^\prime})) .
\end{align}
Here, $M$ is the mass of the heavy quark and is considered the largest relevant energy scale.
Furthermore, $\bs{p}, \bs{p^\prime}$ are the incoming and outgoing heavy quark momenta, $\bs{k} , \bs{k^\prime}$ are the incoming and outgoing gluon momenta, $\bs{q}$ is the transferred momentum and $q_i^2 \in \left\{q_z^2, q_T^2=q_x^2=q_y^2\right\}$ denotes the longitudinal or transverse momentum transfer, for the (longitudinal) $\kappa_z$ or (transverse) $\kappa_T$ diffusion coefficient, respectively. In this convention, one has $3 \kappa = 2 \kappa_T + \kappa_z$ as well as $\kappa_T / \kappa_z=1 $ for an isotropic system. The integration measures are given by $\int_{\bs{k}} =  \int \frac{ \der k^3}{2 k^0 \left(2 \pi \right)^3}$, where $k^0=M$ for the heavy quark and $k^0 = \left| \bs{k} \right|$ for gluons. 

To leading order in the coupling and the inverse heavy quark mass, the dominant contribution to $\kappa$ is given by t-channel gluon exchange \cite{Moore:2004tg} with the corresponding matrix element in an isotropic approximation
\begin{align}
\label{eq:MatrixElement}
\left| \mathcal{M}_g \right|^2 = \left[N_c C_H g^4 \right] 16 M^2 \,\frac{k_0^2 (1+ \cos^2 \theta_{\bs{k} \bs{k^\prime} } )}{\left(q^2 + m_D^2 \right)^2}\,.
\end{align}
The Debye screening mass is computed as $m_D^2 = 4 \lambda \int \frac{\der^3 p }{(2 \pi)^3} \frac{f(p)}{p}$.

For a more detailed discussion on $\kappa$ we refer the reader to \cite{Moore:2004tg} and our previous work \cite{Boguslavski:2020tqz}. The implementation details are provided in Ref.~\cite{Boguslavski:2023fdm}.

\subsection{Jet quenching parameter}
The transverse momentum broadening of jets is quantified by the jet quenching parameter $\qhat$, which, at leading order, can be calculated in a gluonic plasma in a similar way (see \cite{Boguslavski:2023waw, Boguslavski:2023alu} for implementation details),
\begin{align}
\hat q^{ij} &= \lim_{p\to\infty}\int_{\substack{\bs k\bs k'\bs p'\\q_\perp < \Lambda_\perp}} q_\perp^i q_\perp^j (2\pi)^4\delta^4(P+K-P'-K') \nonumber \\
&\qquad\times \frac{1}{4 (\nc^2-1)}\,\frac{\left|\mathcal M_{gg}^{gg}\right|^2}{p} f(\bs k)\left(1+ f(\bs k')\right).
\label{eq:qhat_general}
\end{align}
Here, $P$, $P'$, $K$ and $K'$ denote the lightlike 4-vectors of the in- and outgoing jet particles as well as in- and outgoing plasma particles, respectively.

Since we consider jets of high energy, the momentum transfer cutoff $\Lambda_\perp$ in the integral should be large compared to the typical momentum $\qs$ (defined in more detail below) \cite{Arnold:2008vd}. To address the effects of the anisotropy of the plasma, we define $\qhat^{ij}$ for different directions. The diagonal components sum to the usual jet quenching parameter $\qhat = \qhat^{yy}+\qhat^{zz}$, for a jet moving in the $x$-direction where $z$ is the beam axis.

We use the vacuum gluon scattering matrix element $\mathcal M^{gg}_{gg}$ in the $p\to\infty$ limit with medium effects incorporated via the inclusion of the isotropic HTL self-energy, as explained in more detail in Refs.~\cite{Arnold:2002zm, Boguslavski:2023waw}.

\subsection{Initial conditions}

Our initial conditions are motivated by the highly occupied anisotropic early stages typically encountered in heavy-ion collisions and coincide with \res \cite{Kurkela:2015qoa, Boguslavski:2023alu, Boguslavski:2023fdm}:
\begin{align}
f(\tau{=} 1/\qs, p_\perp, p_z) = \frac{2}{\lambda} A(\xi)\, \frac{\langle p_T \rangle}{\pxi} \exp{\left( \frac{-2  \pxi^2}{3 \langle p_T \rangle^2 } \right)},
\label{eq:IC}
\end{align}
with $\pxi = \sqrt{p_\perp^2 + (\xi p_z)^2}$. We use two sets of parameters with typical initial momentum $\langle p_T \rangle  = 1.8  Q_s$,
\begin{align}
\xi = 10, ~ A(\xi) = 5.24171~; \quad
\xi = 4, ~ A(\xi) = 2.05335.
\end{align}
Here $\xi$ encodes the initial momentum anisotropy of the system and $A(\xi)$ determines the normalization of the distribution such that the initial energy density is the same for both initial parameters.
In all figures, we use the convention that   $\xi=10$ corresponds to full lines and $\xi=4$ to dashed lines. The typical scale of the hard excitations is given by $Q_s$ corresponding to the gluon saturation scale in high-energy QCD.

\section{Bottom-up evolution and time scales}

A weakly coupled system reaches equilibrium according to the bottom-up hydrodynamization scenario \cite{Baier:2000sb}. The dynamics can be grouped into several stages, which we denote by three markers that we use in all figures to disentangle the evolution of different observables (see \app \ref{app:bottom-up} for more details).
The star marker indicates when the occupation number is $1 / \lambda$. This corresponds to an over-occupied stage and closely coincides with the maximum value of the anisotropy. The circle marker is placed at minimum occupancy, which marks the end of the second stage where the anisotropy is approximately constant. Finally, the third stage involves the radiational break-up of hard gluons, which drives the systems towards a hydrodynamic evolution. At the end of this stage, the system is close to equilibrium and isotropy, indicated by the triangle at $P_T/P_L = 2$.

The thermalization time scale of the bottom-up scenario can be parametrically estimated at weak couplings as \cite{Baier:2000sb}
\begin{align}
\tauT(\lambda) = \alpha_s^{\nicefrac{-13}{5}}/Q_s, 
\label{eq:tauT}
\end{align} 
with the coupling constant $\alpha_s = \lambda /(4\pi \nc)$. This is the relevant time scale for the weak-coupling \textit{bottom-up} limiting attractor.

At late times, the approach to isotropy is governed by first-order viscous hydrodynamics. This motivates the definition of a relaxation time $\tauR$ in terms of the shear viscosity since this is the only transport parameter in first-order conformal hydrodynamics. By construction, the relaxation time determines the isotropization rate of the near-equilibrium system and is given by \cite{Heller:2015dha, Heller:2016rtz, Kurkela:2018oqw, Du:2022bel} 
\begin{align}
\tauR(\lambda, \tau) =  \dfrac{4 \pi \,\nicefrac{\eta}{s}(\lambda)}{\Teps(\tau)} .\label{eq:relaxation_time}
\end{align}
In the literature, this time scale is typically used when discussing hydrodynamic attractors \cite{Almaalol:2020rnu, Keegan:2015avk}.
It depends on the coupling via the shear viscosity to entropy ratio%
\footnote{For $1 \leq \lambda \leq 10$ we use the values of $\eta/s$ from Ref.~\cite{Keegan:2015avk}. For $\lambda=0.5$ and $\lambda=20$ we have extracted the $\eta/s$ values, similarly as in Ref.~\cite{Kurkela:2018vqr}, from the late-time behavior of the pressure anisotropy
\begin{align}
    \frac{\eta}{s}(\lambda{=}0.5) = 80\,, \qquad \frac{\eta}{s}(\lambda{=}20) = 0.22\,.
\end{align}
The latter value is consistent with \cite{Kurkela:2018vqr}.}
and on time using the time-dependent effective temperature $\Teps$, which is determined via Landau matching from the energy density, i.e., $\nu_g \frac{\pi^2}{30}\Teps^4(\tau) = \varepsilon$. This is the relevant time scale for the hydrodynamic limiting attractor. A discussion and comparison of the two time scales is provided in \app \ref{app:time_scales}.


\begin{figure}
    \centering
    \includegraphics[width=\linewidth]{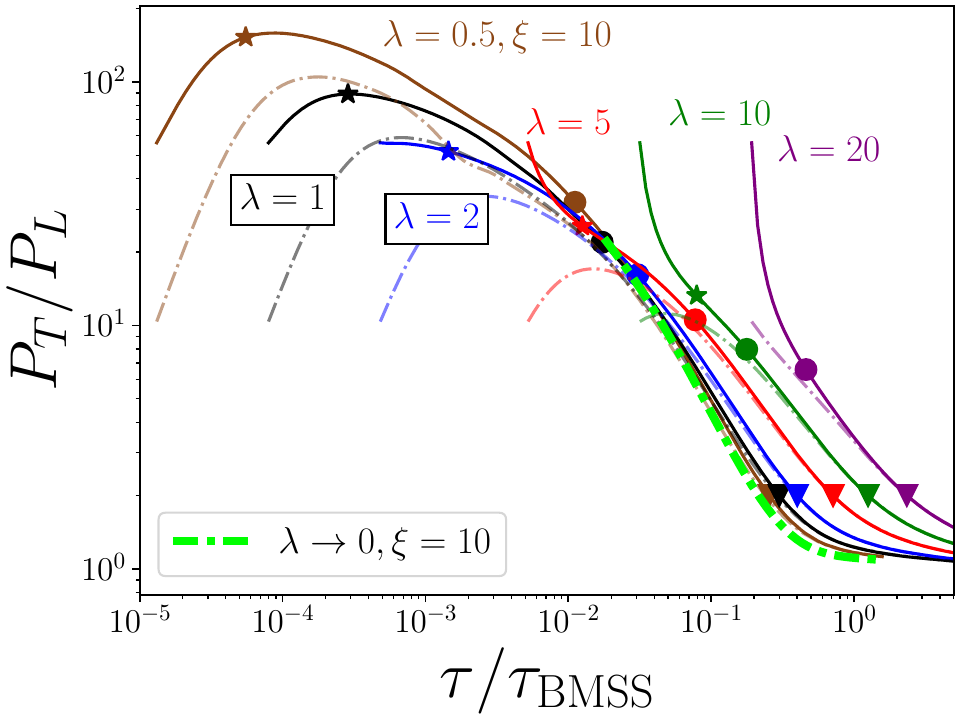}
    \includegraphics[width=\linewidth]{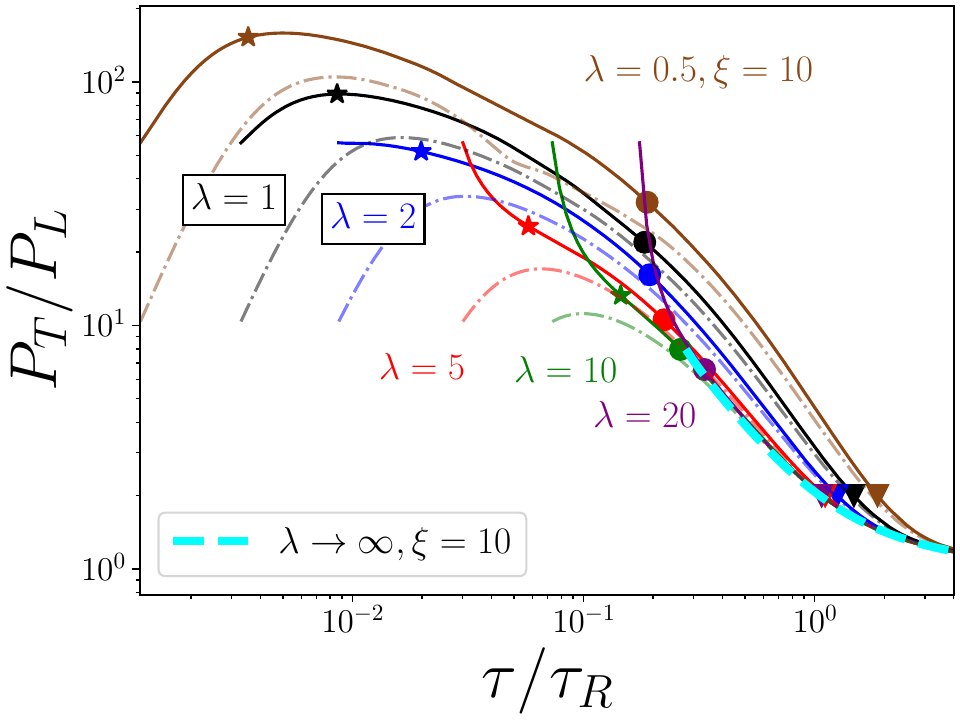}
    \caption{Pressure ratio as functions of $\tau/\tauT$ (top) and $\tau/\tauR$ (bottom), respectively. 
    The extrapolations to vanishing and infinite coupling (the limiting attractors) are performed for each value of $\tau$ as demonstrated in \fig \ref{fig:PTPLratioAndFits}, and are denoted as thick dashed lines.
    }
    \label{fig:pressureRatioVsToverTauR}
\end{figure}

\begin{figure}
    \centering
    \includegraphics[width=\linewidth]{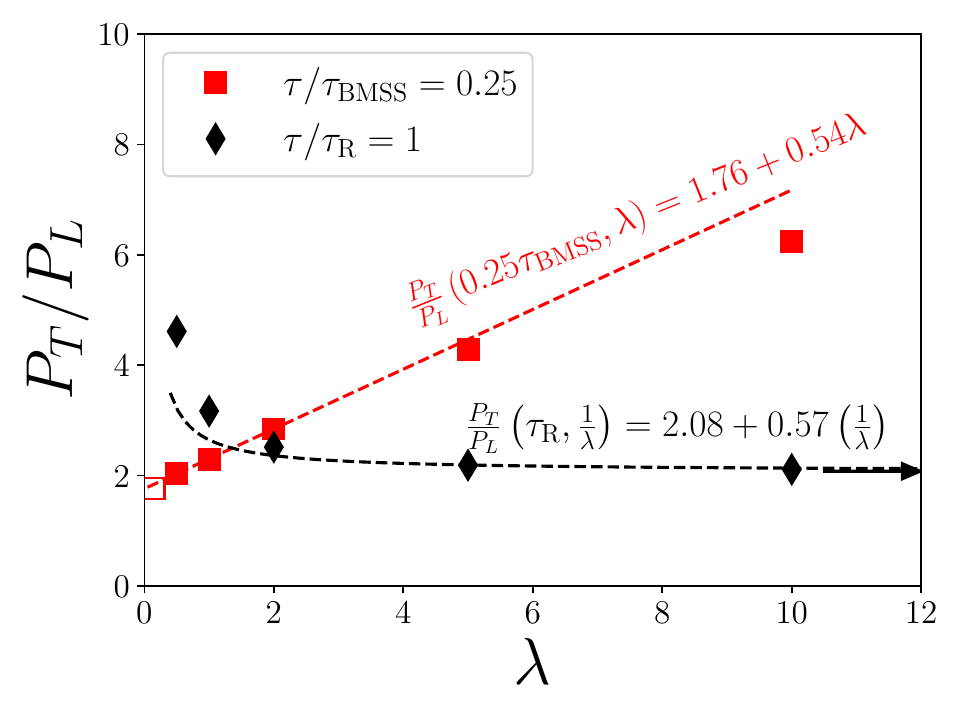}
    \caption{Pressure ratio at a fixed time as a function of the coupling $\lambda$ in the units of both time scales. We also show the fits and their parametrizations, which are used to extrapolate to vanishing coupling (empty square) or infinite coupling (black arrow).
    }
    \label{fig:PTPLratioAndFits}
\end{figure}


\section{Results}
\label{sec:results} 


\begin{figure*}[t]
\centerline{
    \includegraphics[width=0.47\linewidth]{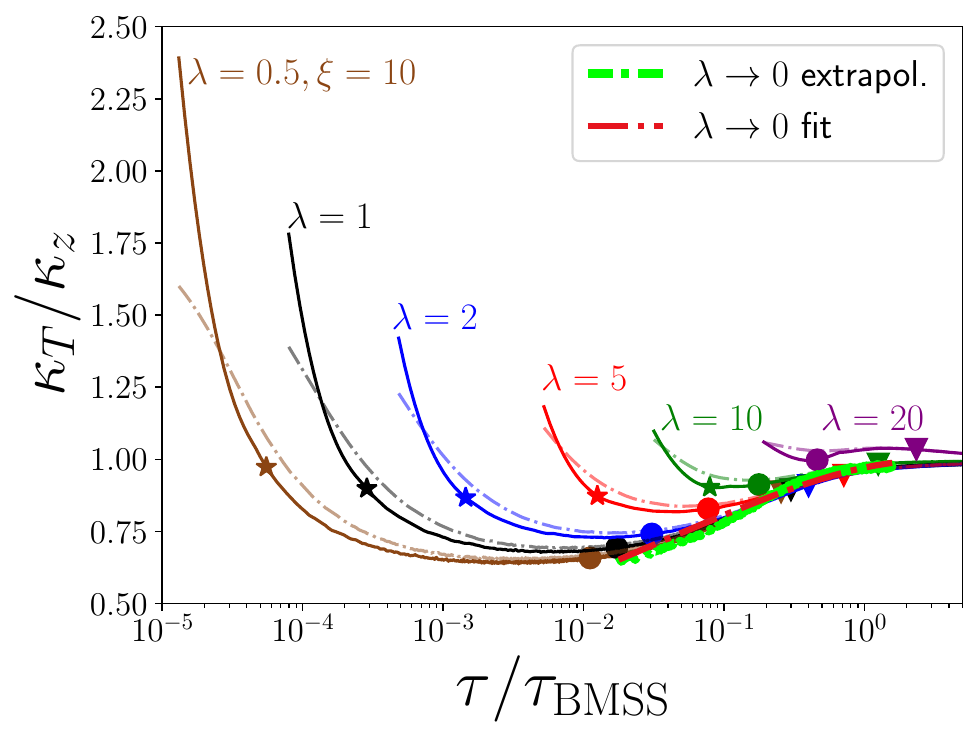}
    \includegraphics[width=0.47\linewidth]{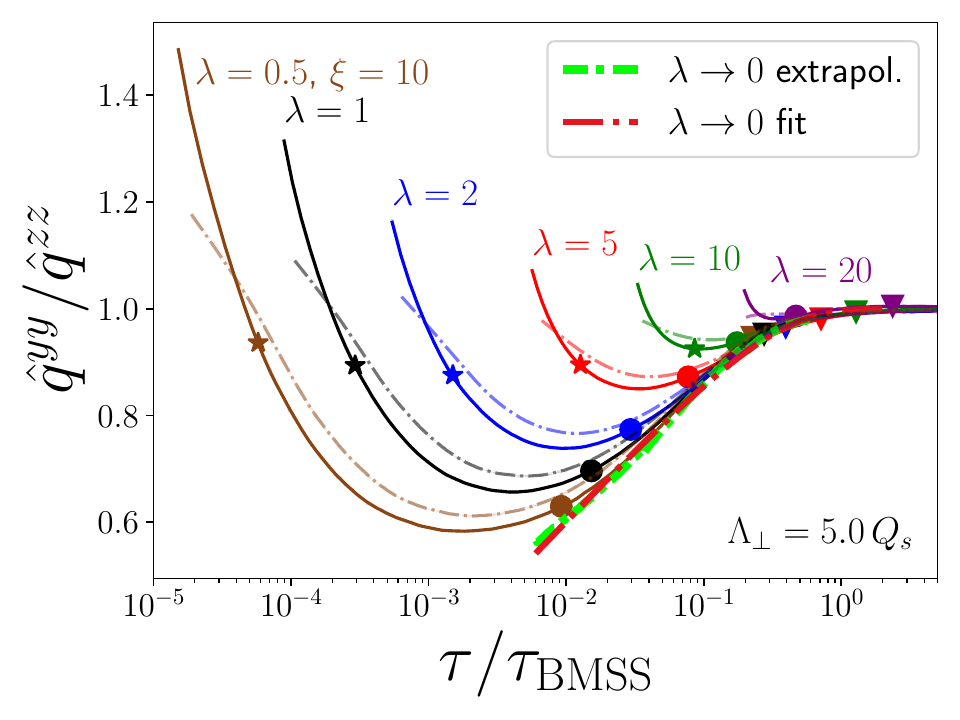}
}
\centerline{
    \includegraphics[width=0.47\linewidth]{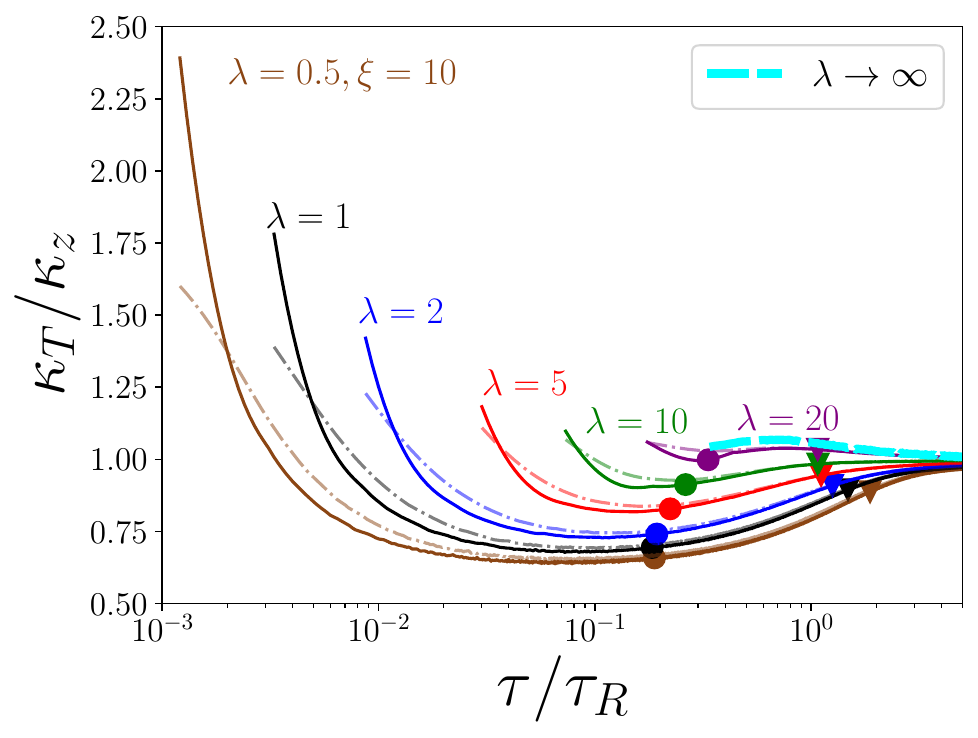}
  \includegraphics[width=0.47\linewidth]{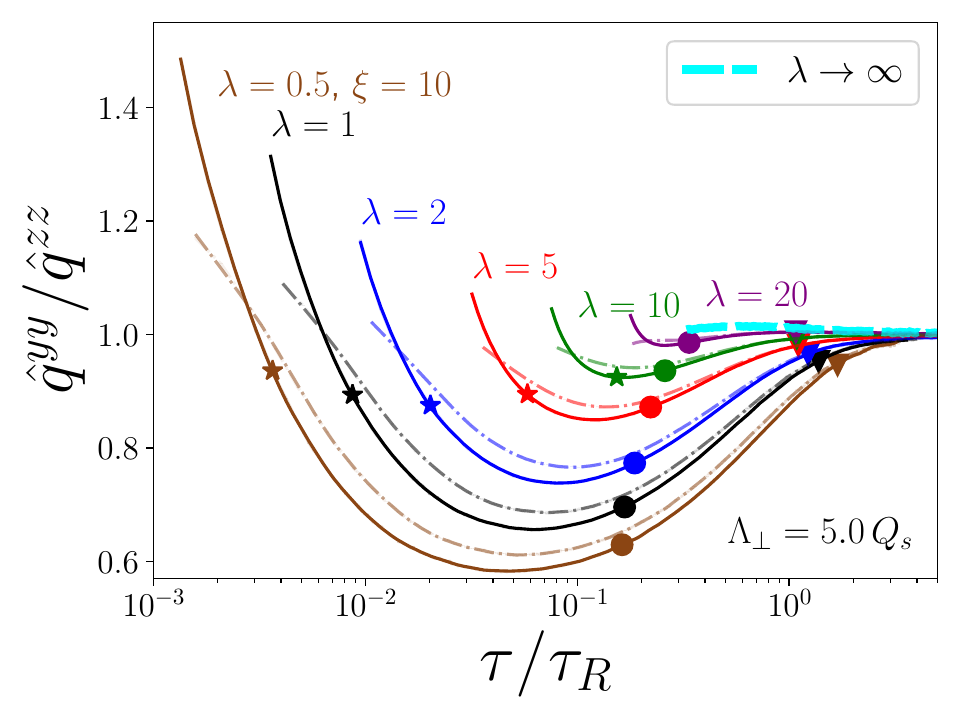}
}
\centerline{
    \includegraphics[width=0.47\linewidth]{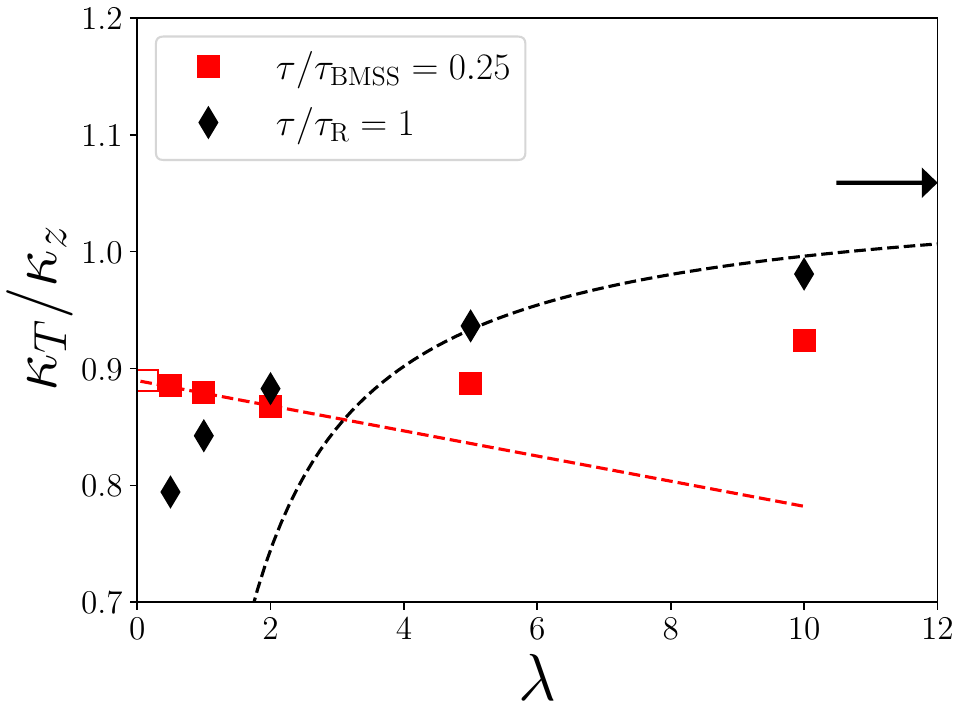}
  \includegraphics[width=0.485\linewidth]{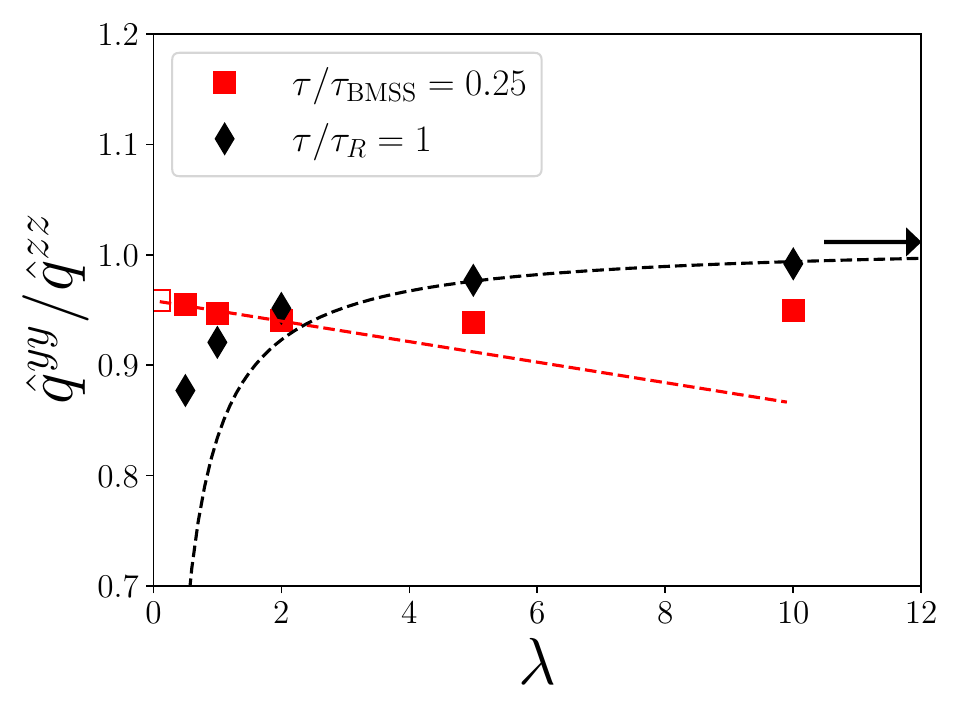}
}

\caption{Ratios of transverse and longitudinal diffusion coefficients using different time scalings, with full and dashed lines corresponding to $\xi=10$ and $\xi=4$, respectively. The left column shows the heavy quark diffusion coefficient ratio $\kappa_T/\kappa_z$ where we have applied a Savitzky-Golay filter to the curves to smoothen the data. 
Similarly, the right column illustrates the ratio of jet quenching parameters $\qhat^{yy}/\qhat^{zz}$ for a fixed $q_\perp$ cutoff $\Lambda_\perp=5\,Q_s$.
The top row shows the quantities as a function of the $\tauT$-time, and the center row as a function of the $\tauR$-time. The bottom row shows the ratios at a specific rescaled time. It illustrates our extrapolation procedure to vanishing coupling (empty square) or infinite coupling (black arrow), with the latter 
performed on the three largest couplings including $\lambda=20$ not shown in the plots. 
\label{fig:kappa_qhat_combined}
}

\end{figure*}


\subsection{Limiting attractors in the pressure ratio}
\label{sec:occupnumberpressratio}

Figure \ref{fig:pressureRatioVsToverTauR} shows the pressure ratio in units of both time scales: the top panel as a function of $\tau/\tauT$
and the bottom panel as a function of $\tau/\tauR$.
One observes that for each coupling, the curves from different initial conditions with $\xi = 4$ and $10$ approach each other, which indicates an attractor behavior.
More systematic studies of the approach to such attractors have been conducted elsewhere \cite{Heller:2015dha, Almaalol:2020rnu, Du:2022bel}.
As observed previously in the literature, some of these attractors resemble each other even far from equilibrium irrespective of different values of the coupling constant%
\footnote{It has been noted that these attractors are similar
even when they are based on different models and degrees of freedom \cite{Keegan:2015avk, Kurkela:2019set}.} 
\cite{Keegan:2015avk, Kurkela:2018vqr, Kurkela:2019set}. For sufficiently large values of the coupling, even the attractors themselves start to overlap. This is visible in the lower panel of \fig\ref{fig:pressureRatioVsToverTauR} and signals additional universal behavior.
This motivates us to define a \emph{limiting attractor} for large couplings, which we obtain by the extrapolation $\lambda\to\infty$ and show as a light-blue dashed line. 
How quickly this hydrodynamic limiting attractor is approached, depends on the value of the coupling.
For large values $\lambda\geq 5$, the approach occurs already 
close to the circle markers.
In contrast, curves at weaker couplings $\lambda \leq 2$ reach the hydrodynamic limiting attractor 
at a significantly later time, after the triangle marker.

For weak couplings, therefore, a different time scaling is more convenient: the bottom-up time scale $\tauT$. Using this to rescale the time variable (top panel of \fig\ref{fig:pressureRatioVsToverTauR}), we observe that the attractors at smaller couplings approach another limiting curve, henceforth called the bottom-up limiting attractor. We obtain it by extrapolating the curves to $\lambda\to 0$ (green dash-dotted line). 
The pressure ratio at weak coupling between the circle and the triangle markers is, therefore, better described by the bottom-up limiting attractor than the hydrodynamic limiting attractor.
It should be noted that at late times,
the curves begin to deviate from the bottom-up and converge to the hydrodynamic limiting attractor instead.
However, this only happens very close to isotropy, while for the pre-equilibrium stage of isotropization, when the deviations from isotropy are of order one, the bottom-up limiting attractor is a better description.

The extrapolation procedure that constructs the limiting attractors visible in \fig \ref{fig:pressureRatioVsToverTauR} is illustrated in \fig \ref{fig:PTPLratioAndFits}. 
For each coupling $\lambda$, we plot the value of the pressure ratio $P_T/P_L$ at a fixed rescaled time $\tau/\tauT = 0.25$. 
Performing a linear fit allows us to extrapolate to a finite value of $P_T/P_L$ for $\lambda \to 0$. This procedure
yields the limiting attractor curve at vanishing coupling.
A similar procedure with the extrapolation $\lambda\to\infty$ via a linear fit in $1/\lambda$ leads to the hydrodynamic limiting attractor.

Thus, we have found two limiting attractors for the pressure ratio, one for $\lambda \to 0$ associated with bottom-up dynamics and one for $\lambda \to \infty$ connected to a viscous hydrodynamic description.

\subsection{Limiting attractors for transport coefficients}
\label{subsec:diffusionquenching}

We now move on to study the  limiting attractors for the jet quenching parameter $\qhat$ and the heavy quark momentum diffusion coefficient $\kappa$. In particular, we focus on the anisotropy ratios of these quantities $\qhat^{yy} / \qhat^{zz}$ for a large transverse momentum cutoff in \eq \eqref{eq:qhat_general} and $\kappa_T/\kappa_z$ in \eq \eqref{eq:kappa_master_formula}. 

These ratios are shown in \fig \ref{fig:kappa_qhat_combined}, with $\kappa_T/\kappa_z$ in the left and $\qhat^{yy} / \qhat^{zz}$ in the right column for a wide range of couplings $\lambda = 0.5$ to $20$. The top row depicts them as functions of time scaled by $\tauT$. One observes a remarkable qualitative similarity in the evolution of both anisotropy ratios. The resulting curves for different couplings and initial conditions are seen to quickly approach each other after the circle marker. This indicates the emergence of universal dynamics already at the far-from-equilibrium state of minimal occupancy. Similarly to the pressure ratio discussed above, this allows us to extrapolate the curves to a bottom-up limiting attractor (light green curve). The linear extrapolation procedure is illustrated in the bottom row at the example of a fixed $\tau/\tauT=0.25$.
We also find that the weaker the coupling, the earlier the anisotropy ratio observables approach the limiting attractor, as can be seen in the upper panels. 
This emphasizes the far-from-equilibrium nature of the bottom-up limiting attractor.

To provide a convenient empirical parametrization, we fit these bottom-up limiting attractor curves to
\begin{align}
    \label{eq:fit1}
    R_{\qhat, \kappa}(\tau) = 1 + c_1^{\qhat,\kappa}\ln\left(1 - e^{-c_2^{\qhat,\kappa} \tau/\tauT}\right).
\end{align}
For the jet quenching ratio $R_{\qhat}(\tau) \approx \qhat^{yy}/\qhat^{zz}$ we obtain $c_1^{\qhat}=0.12$ and $c_2^{\qhat}=3.45$ while for the ratio of the heavy quark diffusion coefficient $R_{\kappa}(\tau) \approx \kappa_T/\kappa_z$ we extract $c_1^{\kappa}=0.093$ and $c_2^{\kappa}=1.33$. 
We include the fits \eqref{eq:fit1} for both anisotropy ratios in the top panels of \fig \ref{fig:kappa_qhat_combined} as dash-dotted lines, labeled `$\lambda\to 0$ fit'. 
One observes that the fitted curves $R_{\qhat, \kappa}$ follow 
the extrapolated bottom-up limiting attractors 
after $\tau \gtrsim 0.01\, \tauT$ to reasonable accuracy, providing an approximate parametrization of the respective limiting attractors. 

We note that the parametrization in \eq \eqref{eq:fit1} should be taken with caution. While its advantage 
lies in its simplicity and small number of fit parameters, the approach toward unity may not be captured completely by this simple ad hoc form~\eqref{eq:fit1}. In particular, we observe that it provides a more accurate description for the bottom-up limiting attractor of the $\qhat$ ratio but shows more pronounced deviations for the $\kappa$ ratio.
Moreover, since the functional form 
can become negative at early times while the $\qhat$ and $\kappa$ anisotropies are always positive in kinetic theory, we expect the fits to deviate substantially from the bottom-up limiting attractors at very early times $\tau \ll 0.01\, \tauT$. Irrespective of the exact parametrization, we emphasize that the bottom-up limiting attractors are well defined at early times.

In contrast, while hydrodynamic limiting attractors for these anisotropy ratios can be defined in a similar way as for the pressure ratio, they offer less predictive power.
We show this in the middle row of \fig \ref{fig:kappa_qhat_combined} where the ratios are depicted as functions of time scaled with $\tauR$. The associated hydrodynamic limiting attractors are obtained by extrapolating to $\lambda\to\infty$ at fixed $\tau/\tauR$ (see the lower panels for an illustration). As is shown in the central panels, the resulting attractors (light blue curves) predict a ratio close to unity long before the system reaches isotropy signaled by the triangle marker.
However, the curves for finite couplings begin to overlap with this limiting attractor only at much later times. 
In particular, this happens even after the triangle marker.
This strongly limits the applicability of the hydrodynamic limiting attractor for these anisotropy ratios. We, therefore, emphasize that the bottom-up limiting attractors provide a much more accurate description of these ratio observables for modeling the pre-equilibrium behavior of hard probes.

\section{Conclusions}
\label{sec:conc}

In this paper, we have focused on the universal features of a set of observables during the bottom-up thermalization process using QCD kinetic theory for a purely gluonic system. 
We establish the concept of limiting attractors, which can be constructed by extrapolating to vanishing or infinite coupling at fixed rescaled times. For the bottom-up limiting attractor in this weak-coupling limit, we use the characteristic timescale $\tauT$ to rescale the time, while for the hydrodynamic limiting attractor (strong-coupling limit) we employ the relaxation time $\tauR$.

Our main result is that the bottom-up limiting attractor provides a good description for the pre-hydrodynamic evolution of jet and heavy-quark momentum broadening ratios $\qhat^{yy} / \qhat^{zz}$ and $\kappa_T/\kappa_z$
during the late stages of the bottom-up thermalization scenario. We added a parametrization of the respective curves for convenience.
In contrast, for these ratios, the often-used hydrodynamic (limiting) attractor predicts only very small deviations from unity even long before the system reaches isotropy (as measured by the pressure ratio) and thus, does not capture the non-trivial pre-equilibrium dynamics at finite couplings.

We emphasize that these two limiting attractors are not contradictory, but rather complementary. However, their usefulness depends on the considered observable. For the aforementioned momentum broadening ratios, for instance, the bottom-up limiting attractor provides a considerably more useful description at finite couplings.
For the pressure ratio $P_T/P_L$ on the other hand, both attractors provide a good description in their respective coupling regimes.

Our results can be used to study the impact of the initial stages on physical observables such as hard probe measurements. For instance, we have observed that for a wide range of couplings the jet quenching parameter ratio $\qhat^{yy} / \qhat^{zz}$ follows the universal bottom-up limiting attractor, which deviates from unity. 
Such deviations 
have been indeed argued to lead to a possible jet polarisation \cite{Hauksson:2023tze}. 
This effect could be more pronounced in medium-sized systems like oxygen-oxygen collisions since they can be particularly sensitive to the out-of-equilibrium dynamics.
Moreover, the bottom-up limiting attractor of $\kappa_T/\kappa_z$
can be a useful tool for studying the impact of initial anisotropic stages on heavy-quark observables, which provides a promising feature.

An important conclusion from our work is that the applicability of hydrodynamic attractors has to be verified for each observable separately, and in some cases the bottom-up limiting attractor is more relevant for the observable under consideration.
While in this paper we provide concrete examples of such observables, a classification of other relevant observables in terms of the sensitivity to the different limiting attractors is left for future studies.

\begin{acknowledgments}
The authors would like to thank X.~Du, I.~Kolbe, A.~Mazeliauskas, D.~M\"uller, 
and M.~Strickland for discussions. 
This work was supported under the European Union's Horizon 2020 research and innovation by the STRONG-2020 project (grant agreement No.~824093) and by  the European Research Council  under project ERC-2018-ADG-835105 YoctoLHC.  The content of this article does not reflect the official opinion of the European Union
and responsibility for the information and views expressed therein lies entirely with the authors.  This work was funded in part by the Knut and Alice Wallenberg foundation, contract number
2017.0036. T.L has been
supported by the Academy of Finland, by the Centre of
Excellence in Quark Matter (project 346324) and project
321840. K.B.~and F.L.~are supported by the Austrian Science Fund (FWF) under project P 34455, and F.L.~additionally by the Doctoral Program W1252-N27 Particles and Interactions.

The authors wish to acknowledge the CSC – IT Center for Science, Finland, and the Vienna Scientific Cluster (VSC) project 71444 for computational resources. 
They also acknowledge the grants of computer capacity from the Finnish Grid and Cloud Infrastructure (persistent identifier urn:nbn:fi:research-infras-2016072533 ).

\end{acknowledgments}


\appendix

\begin{figure}
    \centering
    \includegraphics[width=\linewidth]{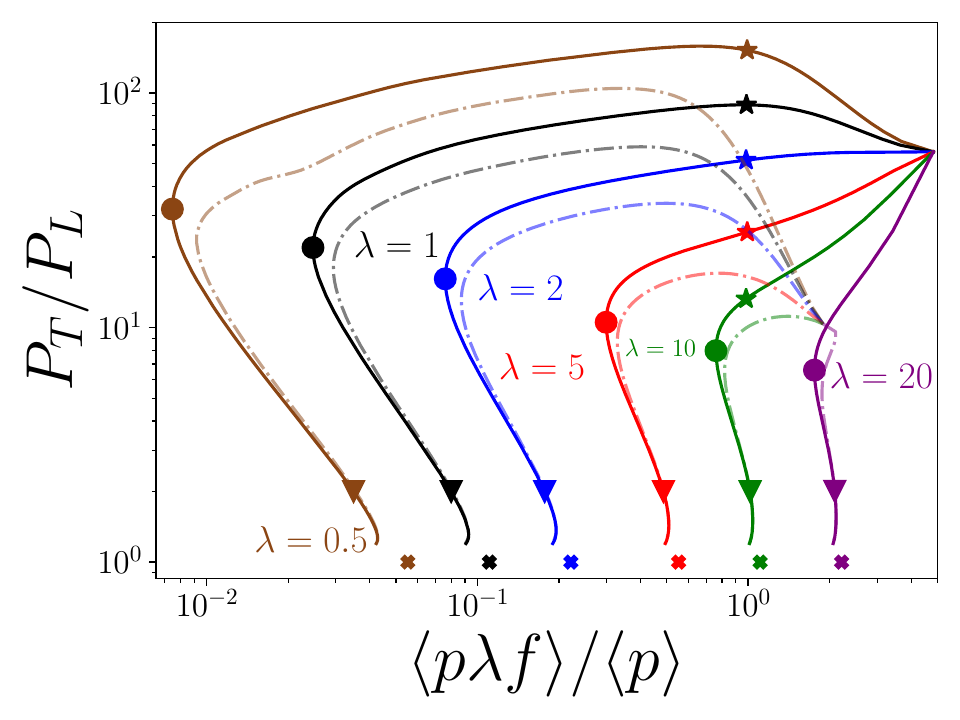}
    \caption{Simulation runs of the thermalization process for various couplings $\lambda$ in the anisotropy-occupancy plane as in Ref.~\cite{Kurkela:2015qoa}. The markers indicate different stages of thermalization as described in the text. \fig \ref{fig:occupVsAnisotropy}. Full lines correspond to $\xi=10$ and dashed lines to $\xi=4$ initial condition. 
    We show the markers only for $\xi=10$ initial condition for clarity.
    }
    \label{fig:occupVsAnisotropy}
\end{figure}

\section{Bottom-up evolution}
\label{app:bottom-up}

The evolution for different couplings and initial conditions is shown in \fig \ref{fig:occupVsAnisotropy} at the example of the pressure ratio $P_T/P_L$ vs.~the typical occupancy of hard particles given by \eq \eqref{eq:occupancy} as in \cite{Kurkela:2015qoa}. The figure displays how an originally highly occupied and anisotropic system evolves towards thermal equilibrium. For weakly coupled systems, the first stage of the bottom-up scenario involves an interplay between the longitudinal expansion and interactions, which
increases the anisotropy even further. This is followed by a stage where the plasma evolves with a roughly constant anisotropy towards underoccupation (after the star marker). During this evolution, a soft thermal bath of gluons is formed. After this (from the circle marker), hard gluons lose their energy to the thermal bath, which ultimately drives the system to thermal equilibrium. The equilibrium is indicated in \fig \ref{fig:occupVsAnisotropy} by crosses.

\begin{figure}
    \centering
    \includegraphics[width=\linewidth]{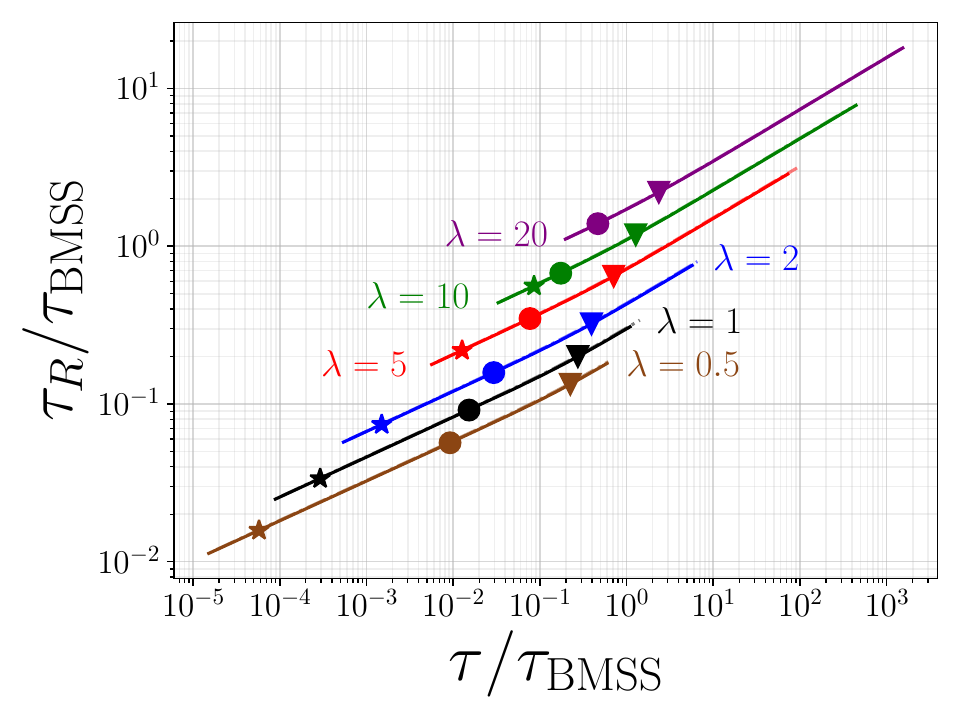}
    \caption{Ratio of the two different time scales $\tauR/\tauT$ during the simulation for different couplings. }
    \label{fig:time_comparison}
\end{figure}

\section{Comparison of relevant time scales}
\label{app:time_scales}

To compare both time scales $\tauT$ and $\tauR$ in \eqs \eqref{eq:tauT} and \eqref{eq:relaxation_time} directly, we show their ratio in \fig\ref{fig:time_comparison}. While the bottom-up time scale $\tauT$ depends only on the coupling $\lambda$ and is therefore constant in time, the kinetic relaxation time $\tauR$
includes also the effective (Landau-matched) temperature, which decreases throughout the time evolution.
For small couplings $\lambda \lesssim 2$, the kinetic relaxation time $\tauR$ is much smaller than the bottom-up thermalization estimate $\tauT$ for our entire simulation, as visible in \fig\ref{fig:time_comparison}. In contrast, for larger values of $\lambda$, the relaxation time is comparable to and even becomes larger than the bottom-up estimate. In particular, both time scales are approximately identical at the triangle marker ($P_T/P_L=2$) for $\lambda = 10$.

For weakly coupled systems, the observation that $\tauT\gg\tauR$ is consistent with the fact that the bottom-up picture dominates the equilibration process, which can also be seen by the emergence of the bottom-up limiting attractor for small couplings. On the other hand, for larger couplings, we have $\tauR\gtrsim\tauT$, which is in line with the hydrodynamic limiting attractor becoming more dominant for larger couplings.
This provides a simple explanation for the observed behavior in the main text of this paper.

\bibliography{spires}

\end{document}